\documentclass[twocolumn,aps,showpacs]{revtex4}
\usepackage{amsmath}
\usepackage{amssymb}
\usepackage{epsf}

\begin{document}

\title{Quantum kinetic equation and universal conductance fluctuations in graphene}
\author{K. Kechedzhi$^{1}$, O. Kashuba$^{1}$, and Vladimir I. Fal'ko$^{1,2}$}
\affiliation{$^{1}$Department of Physics, Lancaster University,
Lancaster, LA1 4YB,~UK \\
$^{2}$ Laboratoire de Physique des Solides, Universite Paris-Sud,
CNRS UMR 8502, F-91405 Orsay, France}

\begin{abstract}
We analyze universal conductance fluctuations (UCF) in graphene in
the framework of diagrammatic perturbation theory in the metallic
regime. It is shown that strong inter-valley scattering lifts the
valley degeneracy of electronic states, whereas at weak inter-valley
scattering two valleys contribute independently such that the
variance of UCF would be expected to show sample- and
geometry-dependent behavior.
\end{abstract}


\maketitle

The unusual chiral properties of charge carriers in graphene
\cite{Novoselov,QHEbi,Geim,Beenakker} have recently received a lot
of attention. Several theories have been developed
\cite{Aleiner,monoMcCann,WLbi,Titov,WLAndo} interpreting observation
of quantum interference effects in graphene, such as weak
localization magnetoresistance \cite{monoExp,exeter} and the
Josephson proximity effect in superconductor-graphene-superconductor
junctions \cite{Morpurgo}. Low-temperature magnetoresistance
measurements \cite{monoExp,exeterUCF,exeterLPhi} have shown
universal conductance fluctuations (UCF) which appear to be robust
over a wide range of electron concentrations and magnetic fields,
and numerical simulations of transport in monolayer graphene with
charged disorder showed sample-to-sample variation of conductance
\cite{beenakker}. In this Communication we analyze UCF using the
same framework as the earlier weak localization studies
\cite{monoMcCann,WLbi}. Specifically, we study UCF in graphene with
various types of disorder in the fully developed metallic regime
($k_Fl \gg 1$), using quantum kinetic equation for diffusive
transport in graphene and technique of semiclassical Keldysh
functions.

The transport in graphene is determined by the low-energy properties
of charge carriers in the vicinity of corners ($K$-points) of
hexagonal Brillouin zone \cite{kpoints}, called valleys. In the case
of monolayer graphene, this can be described using the Hamiltonian
\cite{wallace,AndoReview},
    \begin{gather}
    {\hat{H}}=v\,\vec{\Sigma}\mathbf{p}+{\hat{h}}_{\mathrm{w}}(\mathbf{p})+\hat{V}(\mathbf{r}),
    \label{h1} \\
    {\hat{h}}_{\mathrm{w}}=-\mu \Sigma _{x}(\,\vec{\Sigma}%
    \mathbf{p})\Lambda _{z}\Sigma _{x}(\,\vec{\Sigma}\mathbf{p})\Sigma
    _{x}. \notag
    \end{gather}%
Here, the basis of bi-spinors $%
\Phi =$[$\phi _{\mathbf{K}_{+},A}$, $\phi _{\mathbf{K}_{+},B}$, $\phi _{%
\mathbf{K}_{-,}B}$, $\phi _{\mathbf{K}_{-},A}$] characterizes
electronic amplitudes on two crystalline sublattices of graphene
($A$ and $B$). $\Sigma_{s}$ and $\Lambda_l$, ($l=x,y,z$) are
$4\times4$ matrices in the valley and sublattice spaces
\cite{matrices}, introduced in Ref.\cite{monoMcCann}. The momentum
$\mathbf{p}=p(\cos\varphi,\sin\varphi)$ is defined with respect to
the $\mathbf{K}$-points \cite{kpoints}. The first, 'Dirac' term in
(\ref{h1}) determines an almost linear spectrum $\epsilon=\pm vp$ of
electrons. The trigonal warping term, ${\hat{h}}_{\mathrm{w}}$,
takes into account a slight trigonal asymmetry of the Fermi line of
graphene in one valley (such that
$\epsilon(\mathbf{K},\mathbf{p})\neq\epsilon(\mathbf{K},\mathbf{-p})$),
which will be treated below as a weak perturbation. Due to the
time-reversal symmetry of the system the trigonal warping has
opposite sign in $\mathbf{K}$ and $\mathbf{K'}$ valleys,
$\epsilon(\mathbf{K},\mathbf{p})=\epsilon(\mathbf{K'},\mathbf{-p})$,
which is taken into account by the valley matrix structure of
$\hat{H}$. The time-reversal-symmetric disorder,
\begin{eqnarray*}
 \hat{V}(\mathbf{r})=\mathrm{\hat{I}}u(%
    \mathbf{r}) + \sum \Sigma_s \Lambda_l u_{sl}(\mathbf{r}),
\end{eqnarray*}
consists of the potential $\hat{I}u(\mathbf{r})$ due to remote
Coulomb charges in, or on, the surface of the substrate (the unit
matrix in the valley and sublattice space) and a generic term, which
takes into account all possible symmetry-breaking local
perturbations. The disorder is characterized by $\langle
u_{sl}(\mathbf{r})u_{s'l'}(\mathbf{r'})\rangle =
\delta(\mathbf{r-r'})\delta_{ss'}\delta_{ll'} w_{sl}$, and this
determines the corresponding scattering rates $\tau^{-1}_{sl} =
\delta_{s,s'}\delta_{l,l'} \pi \gamma w_{s,l}/\hbar$ (where $\gamma
= k_F/(2\pi v \hbar)$ is the Fermi density of states). After
averaging over impurity configurations the scattering rates should
preserve rotational symmetry of graphene which means that $\tau_{xl}
= \tau_{yl} \equiv \tau_{\perp l}$ and $\tau_{sx}=\tau_{sy}\equiv
\tau_{s\perp}$. Anticipating a little, two scattering rates,
$\tau^{-1}_{\textrm{z}}= 4\tau^{-1}_{\perp z} + 2\tau^{-1}_{zz} $
and $\tau^{-1}_{\textrm{i}}=~4~
\tau^{-1}_{\perp\perp}~+~2\tau^{-1}_{z\perp}$ describe valley
block-diagonal and valley block-off-diagonal parts of the
symmetry-breaking disorder potential respectively, whereas the total
scattering rate is defined as $\tau^{-1}=\tau_{0}^{-1} +
\tau_{zz}^{-1} + 2 \tau_{\perp z}^{-1} + 2\tau_{z \perp}^{-1} +
4\tau_{\perp\perp}^{-1} $.

\begin{figure}[t]
\centerline{\epsfxsize=0.85\hsize\epsffile{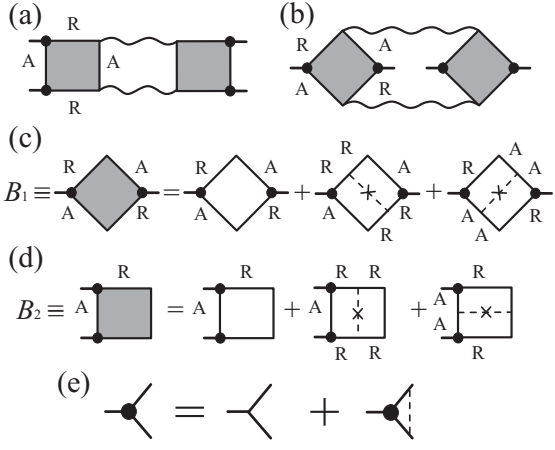}}
\caption{(a),(b) The diagrams which contribute to the main order in
the diagrammatic expansion of the conductivity-conductivity
correlation function. Here the solid lines stand for the impurity
averaged retarded or advanced Green functions, the short wavy tails
stand for the current vertices and the long wavy lines stand for the
diffusion ladders. (c),(d) Hikami boxes of two types and additional
diagrams which determine renormalization in the main order in
$1/k_{F}l \ll 1$. The dashed lines correspond to the disorder
potential. (e) Diagrammatic equation for renormalized current
vertex. } \label{fig:2}
\end{figure}

To characterize the UCF we evaluate the variance of conductance,
$\langle \delta \mathcal{G}^2\rangle = \langle \mathcal{G}^2\rangle
-\langle \mathcal{G}\rangle^2 $, where the angular brackets stand
for averaging with respect to disorder configurations. The main
order of $\langle \delta \mathcal{G}^2\rangle$ in ${1/k_Fl \ll 1}$
is given by perturbation theory diagrams shown in
Fig.\ref{fig:2}(a),(b) \cite{UCF}. These diagrams consist of Hikami
boxes (shaded blocks in Fig.~\ref{fig:2}(a),(b)) connected by the
wavy lines, which represent the sum of ladder diagrams: diffusons
and Cooperons \cite{UCF}. Cooperons are strongly suppressed in
magnetic fields in which magnetic flux is larger then flux quantum
per sample area. Since the UCF are usually studied experimentally in
such a high magnetic field regime, here we neglect the contribution
of the Cooperon diagrams. In contrast, diffusons which are Green
functions of quantum diffusion equation are not suppressed by a
magnetic field although as we will show below their contribution
depends on the efficiency of the symmetry-breaking disorder in the
system. Below, we obtain diffusons by analyzing quantum kinetic
equation in disordered graphene using the semiclassical
approximation ($k_Fl \gg 1$).


\textit{Quantum kinetic equation} describes relaxation of
non-equilibrium inhomogeneous distribution of electrons in a
disordered system. Using matrix Green functions in Keldysh
representation,
$\check{G}(\mathbf{r_1},t_1,\mathbf{r_{1'}},t_{1'})$, we derive
semiclassical form of Green functions for electrons in graphene. We
separate slow and fast variables in $\check{G}$:
$\mathbf{r}=\frac{1}{2}(\mathbf{r_1} + \mathbf{r_{1'}})$ and
$t~=~\frac{1}{2} (t_1~+~t_{1'})$, which vary at $r\gg 1/k_F$ and $t
\gg \hbar/\epsilon_F$, and $\delta\mathbf{r}=
\mathbf{r_1}-\mathbf{r_{1'}}$ and $\delta t~=~(t_1~-~t_{1'})$, which
vary at $\delta r \sim 1/k_F$ and $\delta t \sim \hbar/\epsilon_F$.
We then take the Fourier transform of $\delta G$ with respect to
$\delta t$ and $\delta \mathbf{r}$. The Fourier-transformed Green
function,
    \begin{gather*}
    \check{G}(\epsilon, \mathbf{p}, \mathbf{r}, t) =\left(
        \begin{array}{cc}
        G^{R}(\epsilon, \mathbf{p}, \mathbf{r}, t) & G^K(\epsilon, \mathbf{p}, \mathbf{r},
        t) \\
        0 & G^{A}(\epsilon, \mathbf{p}, \mathbf{r}, t)
        \end{array}\right)
    \end{gather*}
obeys the following Dyson's equation,
    \begin{gather}\label{left}
    \left(\frac{i}{2}
    \partial_{t}
    + \frac{i}{2} v \mathbf{\Sigma
    \nabla_{\mathbf{r}}} - \hat{h}_w  + \epsilon - v\mathbf{\Sigma p}
    - \check{S} \right)\check{G} = 1,
    \\
    \nonumber \check{S} = \int \frac{d^2\mathbf{p}}{(2\pi)^2} \langle\hat{V}
    \check{G}(\mathbf{p}, \epsilon,\mathbf{r},t) \hat{V}\rangle,\label{SE}
    \end{gather}
where $\check{S}$ is the self-energy matrix.

The semiclassical Green function which describes low energy
properties of the system is defined as
\begin{gather*}g(\epsilon,
\mathbf{n}, \mathbf{r}, t) =~\frac{i}{\pi} \int d\xi G( \epsilon,
\mathbf{n}(\epsilon+\xi), \mathbf{r}, t)
\end{gather*}
where integration over $\xi$ is performed in the vicinity of the
Fermi level~\cite{rammersmith}. The response of the system to
external perturbation is described by occupation numbers determined
by the Keldysh component of semiclassical Green function. An
equation for the latter can be derived using the gradient expansion,
that is, assuming that $\check{G} \gg l\nabla_{\mathbf{r}}\check{G},
\tau
\partial_t\check{G}$. Advanced and retarded components of $\check{G}$
are taken into account in the zeroth order of the gradient
expansion,
    \begin{gather*}
    G^{R/A}  =
    \frac{\epsilon+ v \,\mathbf{\Sigma p}}
    {(\epsilon - S^{R/A})^{2} - (vp)^{2}}, \;
    S^{R/A} = \mp\frac{i}{2}\pi\gamma
    \langle\hat{V}\hat{V}\rangle.
    \end{gather*}
This determines the semiclassical Green functions
\begin{gather*}
    g^{R/A} = \pm \frac{1}{2} (\hat{I} + \mathbf{\Sigma n }),
    \end{gather*}
where $\mathbf{n}=\mathbf{p}/p$. Dyson's equation~(\ref{left}) for
Keldysh component of disorder-averaged semiclassical Green function
reads,
    \begin{eqnarray}\label{!!}
    \frac{i}{2}\label{!left}\nonumber
    \partial_{t}g^{K}
    + \frac{i}{2} v \mathbf{\Sigma
    \bigtriangledown_{\mathbf{r}}}g^{K}
    + \epsilon {\Sigma_z} (\hat{I} + \mathbf{\Sigma n }) \Sigma_z
    g^{K} \\
    - \hat{h}_w g^K - S^{R}g^{K} +
    \frac{1}{2}S^{K}(\hat{I} + \mathbf{\Sigma n })
    = 0,
    \end{eqnarray}
where
    \begin{gather*}
    S^{K} = - i\pi \gamma \int \frac{d\theta}{2\pi}
    \langle\hat{V} g^{K} \hat{V}\rangle.
    \end{gather*}
Note that, in Eq.~(\ref{!!}) the energy $\epsilon$ is defined with
respect to mass surface shifted due to effects of disorder and Fermi
line warping \cite{shift}.

Analyzing the main term in Eq.~(\ref{!left}), $\epsilon {\Sigma_z}
(\hat{I} + \mathbf{\Sigma n })\Sigma_z g^{K}$, we find that the
leading contribution to $g^K$ is proportional to the matrix
$(\hat{I} + \mathbf{\Sigma n})$. Using
    \begin{eqnarray}\label{g^K}
    g^{K} = \sum_{i=0,x,y,z}  \left[g^{l} (\hat{I} + \mathbf{\Sigma n}) + \delta\hat{g}^l_z\right]
    \Lambda_l,
    \end{eqnarray}
where $g^{l}$ are functions of $\mathbf{n}$ and $l=0,x,y,z$, we
considered $\delta\hat{g}^l_z$ as a small correction (with an
arbitrary matrix form \cite{matrform}) and checked that the latter
can be neglected in the leading order of the gradient expansion.

Kinetic equation is obtained from equation (\ref{!!}) by subtracting
its hermitian conjugate. After substituting the self-energies and
$g^k$ in the form (\ref{g^K}) we find that,
\begin{multline}
    \partial_t g^{l} + v\mathbf{n \bigtriangledown} g^{l} +
    \frac{1}{\tau} \left( g^{l} - \langle g^l \rangle_{\varphi}
    - \langle g^l \mathbf{n'}\rangle_{\varphi} \mathbf{n}  \right) \\ \label{QKE}
    +\delta^{l} \langle g^l \rangle_{\varphi} + \eta^l \langle g^l
    \mathbf{n'} \rangle_{\varphi} \mathbf{n} + \sum \Upsilon_{ll'} g^{l'} =
    0.
\end{multline}
The angular brackets $\langle ... \rangle_{\varphi}$ in
Eq.~(\ref{QKE}) and below denote averaging over momentum directions,
and the coefficients $\eta^{l}$ and $\delta^{l}$ are defined as
    \begin{gather*}
    \delta^{0}=0,\; \delta^{z}=8\tau _{\perp\perp}^{-1} + 4\tau^{-1}_{z\perp}, \\
    \delta^{x}=\delta^{y}=4\tau _{\perp\perp}^{-1} + 4\tau^{-1}_{\perp z} + 2\tau^{-1}_{z\perp}
    +2\tau _{zz}^{-1},\\
    \eta^0= 4\tau _{\perp\perp}^{-1}
    + 2\tau _{\perp z}^{-1} + 4\tau _{z\perp }^{-1} + 2\tau^{-1}_{zz},\\
    \eta^z = 4\tau _{\perp\perp}^{-1} + 2\tau^{-1}_{\perp z} +2\tau^{-1}_{z
    z}, \\
    \eta^{x} = \eta^{y}= 4\tau _{\perp\perp}^{-1} + 2\tau^{-1}_{\perp
    z} + 2\tau^{-1}_{ z \perp}.
    \end{gather*}
The effect of the Fermi line asymmetry is taken into account by
    \begin{gather*}
    \Upsilon_{xy}=-\Upsilon_{yx}~=~\frac{\hbar v^2}{2\mu\epsilon^2_F} n_x
    (1-4n_y^2),
     \end{gather*}
whereas $\Upsilon_{ll'}$ with $(l,l')\neq (x,y),(y,x)$ are equal to
zero.

The gradient expansion of Eq.~(\ref{QKE}) leads to the diffusion
equation for the angle-average density matrix $\langle
g^l\rangle_{\varphi}$,
    \begin{gather}\label{diff0}
    \left(\partial_t + D_l (i \bigtriangledown)^2  + \Gamma^l\right) \langle g^l\rangle_{\varphi} = 0,
    \end{gather}
where \cite{conductivity}
    \begin{gather*}
    D_l=v^2\tau^l_{tr}/2, \;\;\; \tau_{tr}^l \equiv
    2\tau/(1+\eta^l\tau),\\
    \Gamma _{0}^{0}=0,\;\Gamma _{0}^{z}=2\tau
    _{\mathrm{i}}^{-1},\;\Gamma
    _{0}^{x}=\Gamma _{0}^{y}=\tau _{\mathrm{w}}^{-1}+\tau _{\mathrm{z}%
    }^{-1}+\tau _{\mathrm{i}}^{-1}\equiv \tau _{\ast }^{-1}.
    \end{gather*}
Here, the valley-dependent transport times $\tau^{l}_{tr}$ are
determined by the efficiency of backscattering of electrons in
corresponding mixed valley states (described by density matrix
components $\langle g^l \rangle,\;(l=0,x,y,z)$). Such that, for
example, due to the chirality of electrons in graphene
backscattering off the potential disorder is
suppressed~\cite{AndoReview,monoMcCann}, and in a sample with purely
potential disorder the transport time is given by
$\tau^{l}_{tr}=2\tau_0$ $(l=0,x,y,z)$. In realistic samples
symmetry-breaking disorder restores the backscattering, which
results in reduced valley-state-dependent transport times
$\tau^l_{tr} <2\tau_0$. The relaxation gaps $\Gamma^l$ are induced
in this picture by symmetry-breaking disorder and the Fermi line
warping effect. This warping effect determines the different
evolution operators of electrons in two different valleys, and
suppresses the inter-valley coherence terms in the density matrix
$\langle g^{x/y}\rangle$. At the same time, however, the
intra-valley components of density matrix $\langle g^{0/z} \rangle$
are not affected by the trigonal warping. This effect is taken into
account using a relaxation time,
\begin{gather*}
\tau _{\mathrm{w}}^{-1}=2\tau _{0}\left( \epsilon ^{2}\mu /\hbar
v^{2}\right) ^{2}
\end{gather*}
(we assumed $\tau _{\mathrm{w}}^{-1} \ll \tau^{-1}$). Similarly,
disorder terms $u_{0s}\Lambda_{0}\Sigma_s$ and
$u_{zs}\Lambda_{z}\Sigma_s$, $s=0,x,y,z$ scatter electrons in
different valley states differently, which also leads to relaxation
of $\langle g^{x/y}\rangle$ without affecting $\langle
g^{0/z}\rangle$. Finally, inter-valley disorder terms
$u_{xs}\Lambda_x\Sigma_s$ and $u_{ys}\Lambda_y\Sigma_s$ mix the two
valley states and lead to relaxation of all "valley-triplet"
components of density matrix $\langle g^{x/y/z}\rangle$, which is
taken into account in Eq.~(\ref{diff0}) by the inter-valley
relaxation rate $\tau _{\mathrm{i}}^{-1}$.

\textit{Diffusons}, $\mathcal{D}^{l}$ can be now found as Green
functions of diffusion equations (\ref{diff0}) with initial
inhomogeneous distribution $g^l_0$. To describe UCF in a small
graphene sample we solve diffuson equations (\ref{diff0}) with
boundary conditions at current contacts $\mathcal{D}^{l} = 0$. The
physical edge of graphene is atomically sharp and hence generates
strong inter-valley scattering, which suppresses valley-triplet
diffuson modes near the edge thus leading to the boundary condition,
$\mathcal{D}^{x/y/z}=0$. In contrast, the particle-density
("singlet") mode $\mathcal{D}^0$ has boundary condition
${\mathbf{(n\nabla)}\mathcal{D}^0 = 0}$ corresponding to the absence
of charge current through the edge. Solutions of the diffusion
equations (\ref{diff0}) for a rectangular graphene wire $L_x\times
L_y$ are given by,
    \begin{gather}
    \mathcal{D}^{l}(\mathbf{r},\mathbf{r'})=\frac{1}{\pi\nu\tau^2}
    \sum^{\infty}_{n=1}\sum^{\infty}_{m=0}
    \frac{\phi^{\,l}_{n, m}(\mathbf{r})\phi^{\,l}_{n, m}(\mathbf{r'})}{D^l\pi^2\lambda^{l}_{n, m}}, \nonumber \\
    \phi^{0}_{n,m}(\mathbf{r}) = \sqrt{\frac{2}{L_{x}}\frac{2}{L_{y}}}
    \sin\left(\frac{n\pi x}{L_{x}}\right)\cos\left(\frac{m\pi
    y}{L_{y}}\right),\nonumber\\\nonumber
    \phi^{x/y/z}_{n,m}(\mathbf{r}) = \sqrt{\frac{2}{L_{x}}\frac{2}{L_{y}}}
    \sin\left(\frac{n\pi x}{L_{x}}\right)\sin\left(\frac{m\pi
    y}{L_{y}}\right),\\\label{lambda}
    \lambda^l_{n,m} = \left( \frac{n^2}{L^2_{x}} +
    \frac{m^2}{L^2_{y}}\right) + \frac{\Gamma^{l}+
    \tau^{-1}_{\varphi}}{D^l\pi^2},
    \end{gather}
where we take into account dephasing due to inelastic processes
$\tau_{\varphi}$.

As compared to the conventional electrons systems, \textit{Hikami
boxes} $B^l_1 =
\frac{1}{2}e^2v^2_0\nu\tau^2\frac{(\tau^0_{tr})^2}{\tau^l_{tr}},
B^l_2 = \frac{1}{4}e^2v^2_0\nu\tau^2
    \frac{(\tau^0_{tr})^2}{\tau^l_{tr}}$ and the current vertex ${\widetilde{v}_x
= v_0\tau^0_{tr}/\tau\Sigma_x}$ in monolayer graphene have to be
renormalized by additional diagrams shown in Fig.~\ref{fig:2}(c),(d)
and by vertex corrections (black dots in Fig.~\ref{fig:2}). Both of
these corrections contribute to the variance of conductivity in the
main order in $1/k_Fl$ and are non-vanishing since current operator
in monolayer graphene is momentum independent \cite{monoMcCann}.

\textit{The variance} of conductance fluctuations is a sum of
diagrams shown in Fig.~\ref{fig:2}(a,b), in which the diagram
Fig.~\ref{fig:2}(a) is encountered twice in the diagrammatic
expansion \cite{AKh} and hence has a combinatorial pre-factor $2$.
As a result at $T=0$ we get~\cite{socomment},
    \begin{eqnarray}\label{F(0)}
     \langle \delta \mathcal{G}^2\rangle=\frac{6}{L_{x}^4}\left(\frac{2e^2}{h}\right)^2 \sum_{l,n,m}
    \frac{C_l}{\left[\lambda^l_{n,m}\right]^2},\;\;
    C_l=\left(\frac{\tau_{tr}^0}{\tau_{tr}^{l}}\right)^4.
    \end{eqnarray}

It is interesting to compare UCF for rectangular phase-coherent
graphene samples with $L_x \gg L_y$ and $L_x \ll L_y$. As mentioned
before, we consider the system in an intermediate magnetic field. In
a narrow wire, $\mathcal{D}^{x,y,z}$ decay at the length $\sim L_y$,
and the variance of conductance is dominated by the "valley-singlet"
diffuson component, $\mathcal{D}^{0}$, and,
$\langle\delta\mathcal{G}^2\rangle = {\textstyle \frac{1}{15} \left(
\frac{2e^2}{h} \right)^2}$ \cite{UCFnumerics}, which coincides with
the standard result for quasi 1D metallic wires in the unitary limit
\cite{UCF}. In contrast, in the case of $L_y\gg L_x$, all diffuson
components $\mathcal{D}^{l},l=0,x,y,z$ may contribute to the
variance, depending on whether the effect of trigonal warping
induces suppression of inter-valley diffuson components
$\mathcal{D}^{x/y}$ or not. This determines
$\langle\delta\mathcal{G}^2\rangle = {\textstyle \alpha
\tfrac{3\zeta(3)}{2\pi^3} \tfrac{L_y}{L_x} \left( \frac{2e^2}{h}
\right)^2}$ ($\zeta(n)$ is Riemann's zeta function) with $\alpha=4$
for $L_x < \sqrt{D_0\tau_{\ast}}$, $\alpha=C_0+C_z\approx 2$ for
$\sqrt{D_0\tau_{\ast}} < L_x < \sqrt{D_0\tau_{i}}$, and $\alpha=1$
for $\sqrt{D_0\tau_{\ast,i}}<L_x$.

Inelastic processes such as electron-electron or electron-phonon
interactions limit the coherence length $L_{\varphi} \sim
\sqrt{D_0\tau_{\varphi}} < L_{x/y}$ in the sample. In this case
conductance of the sample is determined as conductance of a network
of resistors of size $L_{\varphi}$ each with conductance variance
given by (\ref{F(0)}). Also, if at high temperatures $ L_T \equiv
\sqrt{\hbar D_0/kT} < L$ and $L_T < L_{\varphi}$ thermal broadening
produces an additional self-averaging reducing the conductance
fluctuations. For a square sample of size $L\times L$: $\langle
\delta \mathcal{G}^2(T)\rangle \sim
    \left(\frac{e^2}{h}\right)^2
    \left(\frac{L_T}{L}\right)^2 \ln\left( \frac{L_{\varphi}}{L_T} \right)$ \cite{UCF}.

In conclusion, we have shown that the variance of
interference-induced conductance fluctuations in graphene is of the
order of the usual UCF value in metals, with a pre-factor dependent
on the strength of the inter-valley scattering and shape of the
sample. In a long wire of graphene or in the material with strong
inter-valley scattering, the magneto-fluctuations of conductance
have the variance typical for the unitary symmetry-class
(intermediate magnetic field). In a wide graphene sample ($L_x <
L_y$) with weak inter-valley scattering, the size of
magneto-conductance fluctuations is increased as compared to unitary
symmetry-class result by at least a factor $\sim 2$. This behavior
is opposite to what was found for the weak localization
magnetoresistance~\cite{WLbi}: the latter was suppressed in the case
of weak inter-valley scattering, whereas strong inter-valley
scattering was found to restore the weak localization effect. This
behavior contrasts the observation that in usual metals with
non-chiral electrons UCF scale similarly to the weak localization
correction to conductivity, made by Aleiner and Blanter \cite{AB}.
The analysis of the UCF in bilayer graphene showed a result very
similar to the monolayer case despite a
difference~\cite{McCann,WLbi} in the electronic spectrum.

The authors wish to thank E.~McCann, B.~Altshuler, A.~Morpurgo,
C.W.J. ~Beenakker, and A.~Savchenko for helpful discussions. This
project has been funded by EPSRC grant EP/C511743 and ESF FoNE
network SpiCo. We also thank K.~Efetov for attracting our attention
to that our UCF results for the intermediate asymptotic regimes
coincide with those in \cite{KhE}.


\begin{thebibliography}{99}
\bibitem{kpoints} Corners of the hexagonal Brillouin zone are $\mathbf{K}_{\pm }
= \pm ({\textstyle\frac{2}{3}}ha^{-1},0)$ ($a$ is the lattice
constant).
\bibitem{matrices} Isospin and pseudospin matrices are defined as,
    \begin{gather*}
    \Sigma_x = \left[
        \begin{array}{cccc}
        0 & 1 & 0 & 0 \\
        1 & 0 & 0 & 0 \\
        0 & 0 & 0 & -1 \\
        0 & 0 & -1 & 0 \\
        \end{array}\right],
    \Sigma_y =
        \left[
        \begin{array}{cccc}
        0 & -i & 0 & 0 \\
        i & 0 & 0 & 0 \\
        0 & 0 & 0 & i \\
        0 & 0 & -i & 0 \\
        \end{array}\right], \\
    \Sigma_z = \left[
        \begin{array}{cccc}
        1 & 0 & 0 & 0 \\
        0 & -1 & 0 & 0 \\
        0 & 0 & 1 & 0 \\
        0 & 0 & 0 & -1 \\
        \end{array}\right],
    \Lambda_x =
        \left[
        \begin{array}{cccc}
        0 & 0 & 1 & 0 \\
        0 & 0 & 0 & -1 \\
        1 & 0 & 0 & 0 \\
        0 & -1 & 0 & 0 \\
        \end{array}\right], \\
    \Lambda_y =
        \left[
        \begin{array}{cccc}
        0 & 0 & -i & 0 \\
        0 & 0 & 0 & i \\
        i & 0 & 0 & 0 \\
        0 & -i & 0 & 0 \\
        \end{array}\right],
    \Lambda_z =
        \left[
        \begin{array}{cccc}
        1 & 0 & 0 & 0 \\
        0 & 1 & 0 & 0 \\
        0 & 0 & -1 & 0 \\
        0 & 0 & 0 & -1 \\
        \end{array}\right].
    \end{gather*}
This matrices determine two commuting subgroups of the group U$_{4}$
of unitary transformations of a 4-component $\Phi $: an isospin
(sublattice) group SU$_{2}^{\Sigma }\equiv \{e^{ia\vec{n}\cdot
\!\vec{\Sigma}}\}$ and a
pseudospin (valley) group SU$_{2}^{\Lambda }\equiv \{\mathrm{e}^{ib\vec{n}%
\cdot \!\vec{\Lambda}}\}$. In this basis time reversal $T(W)$ of an
operator is given by $T(W)=\Lambda_{y}\Sigma_{y}W^{\ast} \Lambda_{y}
\Sigma _{y}$.
\bibitem{shift} Integration of Eq.~(\ref{left}) with respect to $\xi$ gives rise
to terms of the form $\tfrac{i}{\pi} \int d\xi \xi
G(\epsilon,\mathbf{n}(\epsilon + \xi), \mathbf{r},t) = \Delta
\epsilon g(\epsilon,\mathbf{n},\mathbf{r},t)$, where $\Delta
\epsilon$ is a shift of the mass surface due to disorder and Fermi
line warping. Such terms can be eliminated from the equation by
recalculating the energy $\epsilon$ with respect to the shifted mass
surface.
\bibitem{matrform} Any matrix in the $4$D isospin space can be expanded in the basis
of linearly independent matrices  ${[(\hat{I} + \mathbf{\Sigma n}),
(\hat{I} + \mathbf{\Sigma n}) \Sigma_z, \Sigma_z (\hat{I} +
\mathbf{\Sigma n}), \Sigma_z (\hat{I} + \mathbf{\Sigma
n})\Sigma_z]}$. Substitution of the solution in form of the
Eq.~(\ref{g^K}) into the main order in the Eq.~(\ref{!!}) leaves
only terms proportional to the last three matrices in the chosen
basis in~$\delta \hat{g}^l_z$.
\bibitem{conductivity} Conductivity of graphene monolayer with generic disorder is
given by $\sigma = 4 e^2 \nu D_0 = 4 e^2 \nu v^2 \tau^0_{tr}/2$
(where we took into account double spin degeneracy of carriers).
Note that due to the valley-block-diagonal form of the current
operator $v_x = v\Sigma_x$ only $\tau^0_{tr}$ contributes to the
conductance of graphene. At the same time the variance of
conductance contains contribution from mixed valley states and
therefore depends on $\tau^{l}_{tr}, l=0,x,y,z$.
\bibitem{socomment} We take into account
double spin degeneracy of carriers.
\bibitem{UCFnumerics}  This result suggests that the anomalously large variance of
sample-to-sample conductance fluctuations observed in
\cite{beenakker} is caused by a mechanism different from the phase
interference which gives rise to the UCF.
\end{thebibliography}
\end{document}